\documentclass[a4paper,aps,pra,showpacs,twocolumn]{revtex4}

\usepackage{amssymb}
\usepackage{graphicx}
\usepackage{amsmath}
\usepackage{amsbsy}
\usepackage{amsthm}
\usepackage{bbm}
\usepackage{bm}
\usepackage{epsfig}
\newcommand{\id}{\mathbbm{1}}

\newcommand{\be}{\begin{equation}}
\newcommand{\ee}{\end{equation}}
\newcommand{\beq}{\begin{eqnarray}}
\newcommand{\eeq}{\end{eqnarray}}

\usepackage{geometry}
\begin{document}

\title{Experimentally feasible set of criteria detecting genuine multipartite entanglement in n-qubit Dicke states and in higher dimensional systems}

\author{Marcus Huber$^{1}$}
\email{marcus.huber@univie.ac.at}
\author{Paul Erker$^{1}$}
\email{paul.erker@univie.ac.at}
\author{Hans Schimpf$^{1}$}
\author{Andreas Gabriel$^{1}$}
\author{Beatrix Hiesmayr$^{1,2}$}
\affiliation{$^{1}$Faculty of Physics, University of Vienna, Boltzmanngasse 5, 1090 Vienna, Austria}
\affiliation{$^{2}$Research Center for Quantum Information, Institute of Physics, Slovak Academy of Sciences, Dubravska cesta 9, 84511 Bratislava, Slovakia}
\begin{abstract}
We construct a set of criteria detecting genuine multipartite entanglement in arbitrary dimensional multipartite systems. These criteria are optimally suited for detecting multipartite entanglement in $n$-qubit Dicke states with $m$-excitations, as shown in exemplary cases. Furthermore they can be employed to detect multipartite entanglement in different states related to quantum cloning, decoherence free communication and quantum secret sharing. In a detailed analysis we show that the criteria are also more robust to noise than any other criterion known so far, especially with increasing system size. Furthermore it is shown that the number of required local observables scales only polynomially with size, thus making the criteria experimentally feasible.
\end{abstract}


\maketitle
The feature of many body entanglement has recently been recognized as a fundamental property of a broad variety of systems. It appears in quantum phase transitions ({\it e.g.} \cite{phase}) and ionization procedures ({\it e.g.} \cite{helium}). Even biological systems have raised questions as to whether multipartite entanglement might be responsible for their astonishing transport efficiency ({\it e.g.} \cite{Caruso,bio}). There are already numerous examples of possible applications of this quantum feature. In quantum information processing it facilitates quantum computation ({\it e.g.} \cite{qc}), enables multi-party cryptography ({\it e.g.} \cite{SHH1,gisin-crypt}) and also plays a fundamental role in many popular quantum algorithms ({\it e.g.} \cite{brussqa}). \\
In contrast to the well studied bipartite case (for that see {\it e.g.} \cite{horodeckiqe}), the structure of multipartite entanglement is far more intricate. Recently, new tools have been developed to find out whether a given quantum state is multipartite entangled (see {\it e.g.} \cite{horodeckicrit, wocjancrit, yucrit,hassancrit,hhk}). In multipartite systems it is of high importance that criteria detecting multipartite entanglement are experimentally accessible without having to resort to a full state tomography, since that would require an unfeasible number of measurements. For multipartite qubit systems there was some recent progress in locally measurable criteria in Refs.~\cite{seevinckcrit, guehnewit}. These were further developed for two types of multipartite qubit states in Ref.~\cite{guehnecrit}. In multipartite systems many different classes of genuinely multipartite entangled states exist (see {\it e.g.} Refs.~\cite{Wstate,acinbruss,vdmv,Rigolin,HFGSH1,Weinfurter08,HSGSHB1}). The previously introduced criteria only detect a very limited number of such classes to be genuinely multipartite entangled. In Refs.~\cite{HMGH1,GHH1} we have provided a general framework for constructing experimentally feasible criteria in higher dimensional systems. In this letter we introduce a complete set of criteria optimally suited for all n-qubit Dicke states (originally introduced in Ref.~\cite{Dicke}) with an arbitrary number of excitations. In addition we show that these criteria can also be employed to detect different multipartite entangled states related to quantum secret sharing \cite{Weinfurter07}, decoherence free quantum communication \cite{Weinfurter04} and quantum cloning \cite{Murao}.\\

For that purpose let us first briefly review the definition of genuine multipartite entanglement in discrete Hilbert spaces. State vectors $|\psi\rangle$ which can be written as a tensor product
\begin{equation}
|\psi_{A|B}\rangle=|\phi_A\rangle\otimes|\phi_B\rangle\, ,
\end{equation}
with respect to some bipartition $A|B$ are called biseparable. State vectors which are not biseparable with respect to \textit{any} bipartition are genuinely multipartite entangled. This generalizes for mixed states in a straightforward way. Any state $\rho$ which can be decomposed into a convex sum of biseparable pure states 
\begin{equation}
\rho=\sum_i p_i|\psi_{A_i|B_i}\rangle\langle\psi_{A_i|B_i}|\, ,
\end{equation}
is biseparable. Again, any state which is not biseparable is called genuinely multipartite entangled. Note however that the elements of the pure state decomposition need not be biseparable with respect to the same bipartition (i.e. $A_i|B_i\neq A_j|B_j$ is permitted in such decompositions). Therefore a multipartite state can be biseparable, without there existing a bipartition with respect to which it is separable. This gives rise to the difficulty of detecting genuine multipartite entanglement in mixed states. Whereas in the pure state case, it is sufficient to check for biseparability with respect to certain bipartitions ({\it e.g.} with the Peres-Horodecki criterion \cite{horodeckiqe}), this is no longer possible and new tools for detecting entanglement need to be derived.\\
Famous examples of multipartite entangled states are the GHZ (Greenberger Horne Zeilinger) states and the W state, for which the first developed criteria were optimally suited (see Refs.~\cite{guehnecrit,HMGH1} for details). However in multi-body quantum physics there exist a vast and still unknown number of types of multipartite entangled states. One prominent family of multipartite entangled states are the so called Dicke states. They were first studied in the context of coherent light emission (see Ref.~\cite{Dicke}). In recent experiments various different Dicke states were created (see {\it e.g.} Refs.~\cite{Kiesel,Prevedel,Wieczorek,Chiuri}), and different tools used to show that indeed a genuinely multipartite state has been created (see e.g. Ref.~\cite{Toth,Thiel,Campbell,Krammer}). Furthermore Dicke states provide a rich resource for quantum communication tasks. They can be converted into different kinds of resource states (such as GHZ and W) via local operations and classical communication (see Ref.~\cite{Kiesel, Wieczorek})) and can be used for various tasks, such as e.g. open destination teleportation (see Ref.\cite{Prevedel}). Also they exhibit a remarkable robustness of entanglement to particle loss (see e.g. Ref.~\cite{Stockton}).\\
In this letter we provide a simple set of inequalities, which work for all kinds of Dicke states and improve experimental detection, both in terms of noise robustness and feasibility. To that end we first introduce a generalized notion of the originally proposed states for $m$ excitations in $n$ qubit systems, which also incorporates the W state (for the case $m=1$)
\begin{eqnarray}
|D_m^n\rangle=\frac{1}{N}\sum_{\{\alpha\}}|d_{\{\alpha\}}\rangle\,,
\end{eqnarray}
with
\begin{eqnarray}
N={n\choose m}^{-\frac{1}{2}}\,\,\,\text{and}\nonumber\\
|d_{\{\alpha\}}\rangle=\bigotimes_{i\notin\{\alpha\}}|0\rangle_i\bigotimes_{i\in\{\alpha\}}|1\rangle_i
\end{eqnarray}
where $\{\alpha\}$ denotes a set of indices corresponding to the respective subsystems of excitation, and the sum is taken over all inequivalent sets $\{\alpha\}$ satisfying $|\{\alpha\}|=m$.\\
Let us now proceed to construct a set of inequalities which are optimally suited for detecting genuine multipartite entanglement in such states. The set of inequalities
\begin{widetext}
\begin{equation}
\label{mainineq}
\tag{I}
I_m^n[\rho]=
\sum_{\{\gamma\}}\left(\underbrace{|\langle d_{\{\alpha\}}|\rho|d_{\{\beta\}}\rangle|}_{O_{\{\alpha\},\{\beta\}}}-\underbrace{\sqrt{\langle d_{\{\alpha\}}|\otimes \langle d_{\{\beta\}}|\Pi_{\{\alpha\}}\rho^{\otimes 2}\Pi_{\{\alpha\}}|d_{\{\alpha\}} \rangle\otimes |d_{\{\beta\}}\rangle}}_{P_{\{\alpha\},\{\beta\}}}\right)-N_D\sum_{\{\alpha\}}\underbrace{\langle d_{\{\alpha\}}|\rho|d_{\{\alpha\}}}_{D_{\{\alpha\}}}\rangle\leq 0
\end{equation}
\end{widetext}
holds for all biseparable states, where $\Pi_{\{\alpha\}}$ is the cyclic permutation operator acting on the twofold copy Hilbert space (as introduced in Ref.~\cite{HMGH1}), $\{\gamma\}=\{(\{\alpha\},\{\beta\}):|\{\alpha\}\cap\{\beta\}|=m-1\}$, $1<m<\frac{n}{2}$ and $N_D\in\mathbbm{N}$. For every $m$ the inequality is always maximally violated by the corresponding Dicke state $|D_m^n\rangle$, with the number of excitations $m$ equal to $|\{\alpha\}|$, with a value of
\begin{equation}
I_m^n[|D_m^n\rangle\langle D_m^n|]=m\, ,
\end{equation}
For a possible experimental implementation there are two essential issues. One is the resistance to noise and the other is the feasibility in terms of required local measurements. Before we continue to discuss these issues in detail let us quickly prove that ineq.(\ref{mainineq}) indeed holds for all biseparable states. To that end let us examine each off diagonal element $O_{\{\alpha\},\{\beta\}}$ separately for a given bipartition $A|B$
\begin{eqnarray}
O_{\{\alpha\},\{\beta\}}\leq\left\{\begin{array}{cl} P_{\{\alpha\},\{\beta\}}, & \mbox{if }x\in A,y\in B\\ \frac{1}{2}(D_{\{\alpha\}}+D_{\{\beta\}}), & \mbox{if } x,y\in A\vee x,y\in B \end{array}\right.
\end{eqnarray}
where $x=\{\alpha\}\setminus\{\{\alpha\}\cap\{\beta\}\}$ and $y=\{\beta\}\setminus\{\{\alpha\}\cap\{\beta\}\}$. The first statement can be directly proven with inequality (I) from Ref.~\cite{HMGH1} (as $\Pi_{\{\alpha\}}=\Pi_A$ in this case) and the second statement is always true, as it follows from the positivity of the density matrix. Now one just needs to count the number $N_D$ of required $D_{\{\alpha\}}$ elements for the worst case scenario of all possible bipartitions, and the inequality holds. This can be analytically determined in a straightforward way. For $n$ qudits and $m$ excitations
\begin{equation}
N_D=m(n-m-1)\,.
\end{equation}
\begin{figure}[h!]
\begin{center}
\includegraphics[width=5.5cm, keepaspectratio=true]{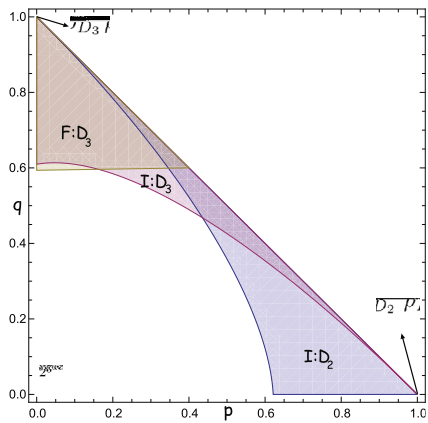}
\caption{(Color online) This plot shows the detection quality of ineq. (\ref{mainineq}) for the six qubit state $\rho=p\rho_{D_2}+q\rho_{D_3}+\frac{1-p-q}{2^6}\id$, where $\rho_{D_2}=|D_2^6\rangle\langle D_2^6|$ and $\rho_{D_3}=|D_3^6\rangle\langle D_3^6|$. Region $I:D_2$ represents the parameter region for which the state is detected to be genuinely multipartite entangled by ineq.(\ref{mainineq}) corresponding to $|D_2^6\rangle$ and analogously $I:D_3$ corresponds in the same way to ineq.(\ref{mainineq}) using $|D_3^6\rangle$. The region $F:D_3$ depicts the parameter region detected by the fidelity witness used in Ref.~\cite{Wieczorek}, which demonstrates the improvement provided by ineq.(\ref{mainineq}).}\label{example}
\end{center}
\end{figure}
Now let us continue to discuss the noise resistance of inequality (\ref{mainineq}), first in a few exemplary cases where comparable criteria exist, and then in the general case. \\
\noindent{\it Four qubits}: For four qubits there exists a comparable criterion, introduced in Ref.~\cite{guehnecrit}. To compare let us consider a four qubit Dicke state with two excitations, mixed with white noise
\begin{eqnarray}
\rho_{ex1}=(1-p)|D_2^4\rangle\langle D_2^4|+p\frac{1}{16}\id\,.
\end{eqnarray}
In this case the criterion from Ref.~\cite{guehnecrit} detects the state as genuinely multipartite entangled up to a noise threshold of $p<\frac{8}{21}\approx0.381$, whereas our criterion works up to a noise threshold of $p<\frac{8}{17}\approx0.471$ and thus has a significantly larger detection range. Recently a new approach based upon semi-definite programming has been proposed in Ref.~\cite{GuehneSDP}. Numerically optimizing over a set of observables they achieve a noise threshold of $p\approx 0.539$, which for four qubits is even better than our proposed criterion. However such a numerical optimization is hardly possible beyond seven qubits. This shows that our criterion is already close to the border of biseparable states, as we achieve comparable thresholds without any numerical optimization involved.\\
For the phased Dicke state introduced in Ref.~\cite{Chiuri} the criterion developed in Ref.~\cite{Krammer} detects the state mixed with white noise to be genuinely multipartite entangled for $p<0.4$, whereas ineq.(\ref{mainineq}) also detects the state up to a noise threshold of $p<\frac{8}{17}\approx0.471$, which also provides a great improvement.\\
For the state originally proposed for quantum telecloning in Ref.~\cite{Murao}, which can be employed also for decoherence free quantum communication \cite{Weinfurter04} and quantum secret sharing \cite{Weinfurter07} mixed with white noise, we can detect genuine multipartite entanglement up to a noise threshold of $p<\frac{16}{43}\approx 0.372$. The original methods used in Ref.\cite{Weinfurter08} can only detect this state up to a noise threshold of $p\approx 0.278$. This is remarkable as it shows that even though the inequalities were designed exploiting combinatorial facts about Dicke states, many other states can also be detected in a very noise robust fashion. Furthermore for this state there exists a feasible experimental proposal in Ref.~\cite{Zukovski}, which has since been realized in three different experiments (see Refs.~\cite{Eibl, Gaertner, Xu}).\\
\noindent{\it Six qubits}: If we proceed to the six qubit case Ref.~\cite{guehnecrit} provides no comparable criterion, but we can compare ineq.(\ref{mainineq}) to fidelity based criteria from Ref.~\cite{Toth}, as done in Fig.~\ref{example}. \\
\noindent{\it $n$ qubits}:
\begin{figure}[h!]
\begin{center}
\includegraphics[width=6cm, keepaspectratio=true]{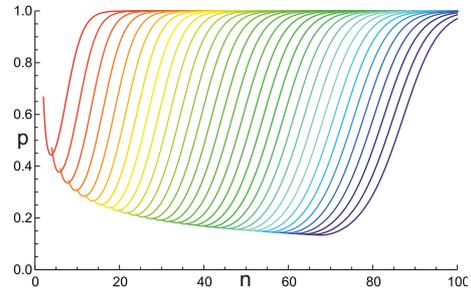}
\caption{(Color online) The plot shows the white noise resistance parameter $p$ in terms of the number of qubits $n$, where from left to right the resistance is plotted for $m=1$ to $m=33$ in ascending order. Notice that after a short dip noise resistance quickly approaches one. Thus the criteria are most suited for detecting $m<\frac{n}{2}$ excitations.}\label{rainbowscaling}
\end{center}
\end{figure}
Let us continue by deriving a general expression for white noise resistance. For $n$ qubit Dicke states with $m$ excitations mixed with white noise, the noise threshold works out as
\begin{equation}
p<\frac{2^n}{2^n+(-1-2 m+2 n)\cdot{n\choose m}}\, ,
\end{equation}
which for the large $n$ quickly approaches $1$ regardless of the chosen $m$. So the white noise tolerance increases rapidly in large systems ({\it e.g.} for 20 qubits and $m=2$ it is already over 99$\%$), which is illustrated in Fig.~\ref{rainbowscaling}.\\
Another crucial issue for experimental implementation comes in terms of feasibility of measurements. With growing system size a full state tomography becomes unachievable, since the number of local measurements required grows exponentially (with $2^{2n}$ for $n$ qubits). The number of density matrix elements which need to be ascertained for ineq.(\ref{mainineq}) grows only polynomially with system size (e.g. for $m=2$ with $n^3$). Also it is clear that any density matrix element can be re-expressed in terms of local expectation values of Pauli operators (as explicitly done in Refs.~\cite{SHH1,HSGSHB1} for similar inequalities). Expressed in terms of local expectation values of Pauli operators, e.g. the inequality for four qubits and $m=2$ reads:
\begin{eqnarray}
I_2^4[\rho]=\sum_{\pi}\sum_{i=x,y}\langle\pi(\id_2\id_2\sigma_i\sigma_i)\rangle-\sum_\pi\sum_{i=x,y}\langle\pi(\sigma_i\sigma_i\sigma_z\sigma_z)\rangle-\nonumber\\(3+3\langle\sigma_z\sigma_z\sigma_z\sigma_z\rangle-\sum_\pi\langle\pi(\id_2\id_2\sigma_z\sigma_z)\rangle)-f(\langle\sigma_z^i\sigma_z^j\sigma_z^k\sigma_z^l\rangle)\nonumber\\
\end{eqnarray}
where $i,j,k,l\in\{0,1\}$, $\pi$ denotes a permutation and $f$ is a simple bilinear function of its arguments. For this example we have replaced $|O_{\{\alpha\},\{\beta\}}|$ by $\Re e[O_{\{\alpha\},\{\beta\}}]$, which does not affect the detection quality for standard Dicke states. So one can explicitly see here that for four qubits there are 39 local measurement settings required, which is a lot more feasible than the 255, required for a full state tomography. The bilinear function $f$ stems from the $P_{\{\alpha\},\{\beta\}}$ term in the inequality and can be straightforwardly be calculated solving the linear equation associating density matrix elements with expectation values. E.g. for $P_{\{0101\},\{0011\}}=\sqrt{\langle0111|\rho|0111\rangle\langle0001|\rho|0001\rangle}$, we can compute
\begin{eqnarray}
\langle0111|\rho|0111\rangle=\frac{1}{8}(1-\langle\sigma_z\sigma_z\sigma_z\sigma_z\rangle-\sum_\pi \langle\id\pi(\id\id\sigma_z)\rangle\nonumber\\
+\langle\sigma_z\pi(\id\sigma_z\sigma_z)\rangle-\langle\id\pi(\id\id\sigma_z)\rangle+\langle\sigma_z\pi(\id\sigma_z\sigma_z)\rangle\nonumber\\
+\langle\sigma_z\id\id\id\rangle-\langle\id\sigma_z\sigma_z\sigma_z\rangle)\, ,
\end{eqnarray}
and of course also all other elements. They all use the same expectation values, but yield a rather cumbersome expression, which is why we chose to abbreviate using $f$.\\
In conclusion the introduced set of criteria are the most general in this field so far. They detect every possible $n$-qubit Dicke state with $m$-excitations to be genuinely multipartite entangled in a very noise robust and experimentally feasible way. In addition they also detect genuine multipartite entanglement in states, used in many practical quantum information processing applications. This provides the opportunity to re-examine already performed experiments as well as designing novel ones. They possibly provide a foundation for experimental classification of multipartite entanglement, as done with similar inequalities in Ref.~\cite{HSGSHB1}.\\
\noindent\emph{Acknowledgements.}
We would like to thank F. Hipp and Ch.Spengler for productive discussions. A. Gabriel and M. Huber gratefully acknowledge the support of the Austrian Fund project FWF-P21947N16.  B.C. Hiesmayr acknowledges EU project QESSENCE.

\end{document}